\documentclass[journal]{IEEEtran}
\usepackage{cite}
\usepackage{graphicx}
\usepackage{textcomp}
\usepackage{amsmath}
\usepackage{algorithmic}
\usepackage{url}

\begin{document}

\title{End-to-End Hardware Modeling and Sensitivity Optimization of Photoacoustic Signal Readout Chains}

\author{Weiran Yang, Yiqi Cai, Handi Deng, and Cheng Ma, \IEEEmembership{Member, IEEE}
\thanks{\emph{Corresponding authors: Cheng Ma; Handi Deng}.}
\thanks{Weiran Yang and Handi Deng are with the Department of Electronic Engineering, Tsinghua University, Beijing 100084, China (e-mail: yangwr22@mails.tsinghua.edu.cn; handi.deng@foxmail.com).}
\thanks{Yiqi Cai was with Beihang University, Beijing, China. He is now with College of Mechanical and Vehicle Engineering, Changsha University of Science \& Technology, Changsha, China (e-mail: caiyiqi@csust.edu.cn).}
\thanks{Cheng Ma is with the Department of Electronic Engineering, Beijing National Research Center for Information Science and Technology, Beijing 100084, China, also with the Institute for Precision Healthcare, Tsinghua University, Beijing 100084, China, and also with the IDG/McGovern Institute of Brain Research, Beijing 100084, China (e-mail: cheng\_ma@tsinghua.edu.cn).}}

\markboth{Yang \MakeLowercase{\textit{et al.}}: End-to-End Hardware Modeling and Sensitivity Optimization of Photoacoustic Signal Readout Chains} {Yang \MakeLowercase{\textit{et al.}}: End-to-End Hardware Modeling and Sensitivity Optimization of Photoacoustic Signal Readout Chains}

\maketitle

\begin{abstract}
The sensitivity of the acoustic detection subsystem in photoacoustic imaging (PAI) critically affects image quality. However, previous studies often focused only on front-end acoustic components or back-end electronic components, overlooking end-to-end coupling among the transducer, cable, and receiver. This work develops a complete analytical model for system-level sensitivity optimization based on the Krimholtz, Leedom, and Matthae (KLM) model. The KLM model is rederived from first principles of linear piezoelectric constitutive equations, 1D wave equations and transmission line theory to clarify its physical basis and applicable conditions. By encapsulating the acoustic components into a controlled voltage source and extending the model to include lumped-parameter representations of cable and receiver, an end-to-end equivalent circuit is established. Analytical expressions for the system transfer functions are derived, revealing the coupling effects among key parameters such as transducer element area (EA), cable length (CL), and receiver impedance (RI). Experimental results validate the model with an average error below 5$\%$. Additionally, a low-frequency tailing phenomenon arising from exceeding the 1D vibration assumption is identified and analyzed, illustrating the importance of understanding the model's applicable conditions and providing a potential pathway for artifact suppression. This work offers a comprehensive framework for optimizing detection sensitivity and improving image fidelity in PAI systems.
\end{abstract}


\section{Introduction}

\IEEEPARstart{P}{hotoacoustic} imaging (PAI) reconstructs tissue structures by detecting ultrasonic waves generated from transient thermoelastic expansion after pulsed laser absorption. The sensitivity of the acoustic detection subsystem is a key determinant of lateral resolution and overall imaging quality \cite{b1}. Conventional PAI detection systems are typically based on piezoelectric transducers, though other transduction mechanisms have been proposed to enhance sensitivity \cite{b2}, \cite{b3}, \cite{b4}, \cite{b5}. However, cost and technical maturity currently limit their application \cite{b6}.

Most prior research on PAI sensitivity focused on either the front-end acoustic acquisition or the transducer itself. Studies have shown that piezoelectric transducers are more responsive to longitudinal than shear waves \cite{b7}, and that detection sensitivity depends on laser fluence, wavelength, imaging depth and the absorption cross-section of the target \cite{b8}, \cite{b9}. Thermal noise remains the fundamental limit to sensitivity, quantified by parameters such as noise equivalent pressure (NEP) and noise equivalent number of molecules (NEM) \cite{b10}. The transducer element area (EA) strongly influences sensitivity-smaller elements suffer reduced signal amplitude \cite{b1}, \cite{b10}, \cite{b11}, \cite{b12}-yet the specific mechanisms by which EA affects sensitivity has not been fully elucidated.

Efforts to improve sensitivity from the back-end circuitry include integrating preamplifiers \cite{b11} and impedance matching networks \cite{b13}, whereas cables reduces sensitivity \cite{b11}. Oraevsky et al. discussed the theoretical ultimate sensitivity with open-circuited and short-circuited transducers, but did not account for cables or general complex receiver impedance (RI) values \cite{b14}. The end-to-end coupling among transducer, cable, and receiver has rarely been investigated systematically, which may trap the design strategy in a local optimum.

A practical method  for analyzing the entire detection system is to represent the piezoelectric transducer as an equivalent electrical circuit, allowing unified treatment with other components. Considering that array transducers consist of single elements, we discusses the case of a single-element piezoelectric transducer receiving a plane wave without loss of generality, which corresponds to a 1D model for the transducer. Among these models, the Krimholtz, Leedom, and Matthaei (KLM) model \cite{b15} is widely used in both research and commercial software like Biosono \cite{b16} due to its high accuracy and physical intuitiveness.

However, the derivation of the KLM model has remained unclear since its introduction \cite{b15}. Theoretical studies have mainly focused on transmitting and receiving coupled transducers in ultrasonics \cite{b17}, \cite{b18}, \cite{b19}, \cite{b20}, \cite{b21}, or transmitting transducers in contexts such as histotripsy \cite{b22}, \cite{b23}, \cite{b24}, \cite{b25}. By contrast, the pure reception process in PAI has received little attention. Nevertheless, the computational effectiveness and simplicity of this model using transfer matrices have been validated \cite{b18}, \cite{b19}, \cite{b20}, \cite{b21}, \cite{b24}. 

In PAI, most studies directly employ the KLM model for simulations without theoretical analysis, such as calculating the input impedance, impulse response (IR), pulse-echo response, and corresponding spectra \cite{b26}, \cite{b27}, \cite{b28}, \cite{b29}, \cite{b30}, \cite{b31}, \cite{b32}, or for transducer designs \cite{b33}, \cite{b34}, \cite{b35}. Lashkari et al. proposed that the ratio of its transmit and receive transfer functions remains constant at any frequency due to reciprocity, allowing the transmit function to simulate normalized detection sensitivity for PAI \cite{b36}, \cite{b37}. While supporting the model’s applicability in PAI in theoretical terms, this research neither clarifies its physical essence nor defines its applicable conditions. As a result, the theoretical gap in the KLM model raises concerns about its specific scalability in PAI and limits its potential applications.

\begin{figure}[!t]
\centerline{\includegraphics[width=\columnwidth]{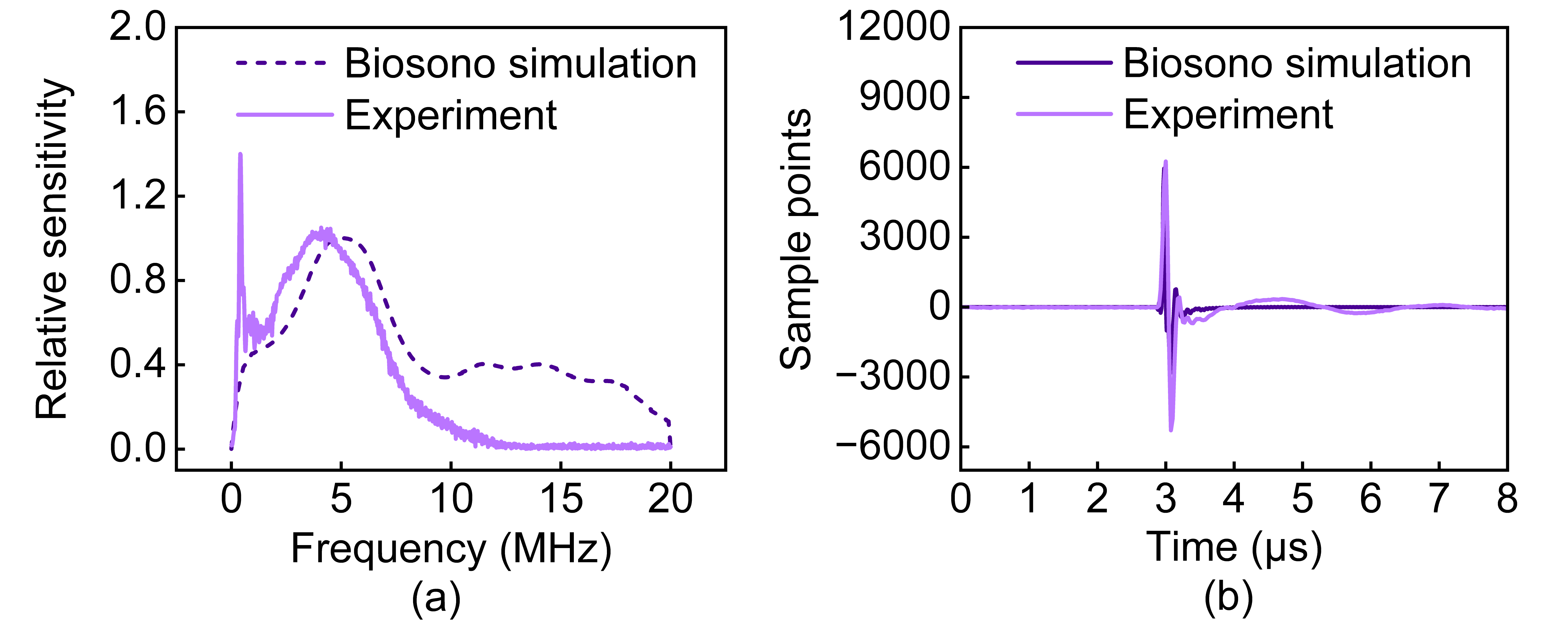}}
\caption{Comparison of Biosono simulation with experimental results from this study: (a) shows the amplitude-frequency response of the detection system, (b) shows its impulse response (IR).}
\label{Fig1}
\end{figure}

Biosono \cite{b16} integrates KLM-modeled transducer, transmission-line-modeled cable, and input resistance of the receiver. Its end-to-end layout not only expands the original KLM model but also reflects the demand for system-level analysis. However, its purely resistive receiver struggles to capture the ubiquitous parasitic reactance, while the distributed-parameter cable significantly complicates parameter input and computation. Furthermore, our experimental validation indicates considerable room for improvement in its accuracy (see Fig.~\ref{Fig1}).

This study rederives the KLM model analytically from linear piezoelectric constitutive equations, 1D wave equations, and transmission line theory, explicitly clarifying its physical essence and applicable conditions. Then its acoustic and electrical components are simplified and expanded respectively to establish an end-to-end equivalent circuit. We replace the distributed-parameter modeling for cable \cite{b16}, \cite{b38} with a lumped T-network and incorporate a complex impedance for receiver. Subsequently, analytical expressions for the transfer functions are derived to characterize relative sensitivity. While maintaining simplicity, the specific mechanisms by which the key parameters — such as EA, cable length (CL), and receiver impedance (RI) — affect sensitivity and their coupling effects are quantitatively analyzed.

Furthermore, distinctive low-frequency peaks that neither Biosono \cite{b16} (see Fig.~\ref{Fig1}(b)) nor our improved model could simulate were observed experimentally in the spectra, they manifest as long tailing signals of IR in time domain and block artifacts in image domain, severely degrading image quality. They were found to stem from a standing wave mode similar to the planar expander-disk (PE-disk) mode \cite{b39} of the piezoelectric plate, namely the radial resonance, extending beyond the 1D vibration assumption. This discovery not only refines the specific applicability conditions of the improved model in PAI but also provides a potential pathway for artifact suppression in PAI systems.

\section{End-to-end PAI acoustic detection system}
\subsection{Theoretical analysis}
\subsubsection{Derivation process, physical essence, and applicable conditions of the KLM model}

For the piezoelectric plate in a single-element transducer, the 1D wave equation can be applied based on the one-dimensional (1D) vibration assumption of the thickness extensional (TE) mode \cite{b39}, and the piezoelectric constitutive equations which are linear under isothermal, adiabatic \cite{b40}, low-frequency \cite{b41}, and small-signal \cite{b42} conditions are introduced. Combining them we yield \cite{b43}:
\begin{equation}
    F_B = \frac{Z_0}{j\tan{kL}}v_B + \frac{Z_0}{j\sin{kL}}v_F + \frac{h_{33}}{j\omega}I.\label{eq1}
\end{equation}
\begin{equation}
    F_F = \frac{Z_0}{j\sin{kL}}v_B + \frac{Z_0}{j\tan{kL}}v_F + \frac{h_{33}}{j\omega}I.\label{eq2}
\end{equation}
\begin{equation}
    V = \frac{h_{33}}{j\omega}v_B + \frac{h_{33}}{j\omega}v_F + \frac{I}{j\omega{C_0}}.\label{eq3}
\end{equation}

$V$ and $I$ represent the voltage and displacement current between the two electrodes, $F_F$, $F_B$, $v_F$, and $v_B$ denote the forces and particle velocities on the front and back surfaces of the plate. $L$, $C_0$, $Z_0$, and $h_{33}$ are its thickness, static clamping capacitance, acoustic impedance, and piezoelectric stress constant respectively. $\omega$ and $k$ denote the angular frequency and wave number under time-harmonic condition. Applying the “impedance analogy” \cite{b44}, we observe that \eqref{eq1}, \eqref{eq2}, and \eqref{eq3} define a three-port linear network featuring two symmetric acoustic ports and a electrical port for which the KLM model provides a specific circuit implementation.

\begin{figure}[!t]
\centerline{\includegraphics[width=\columnwidth]{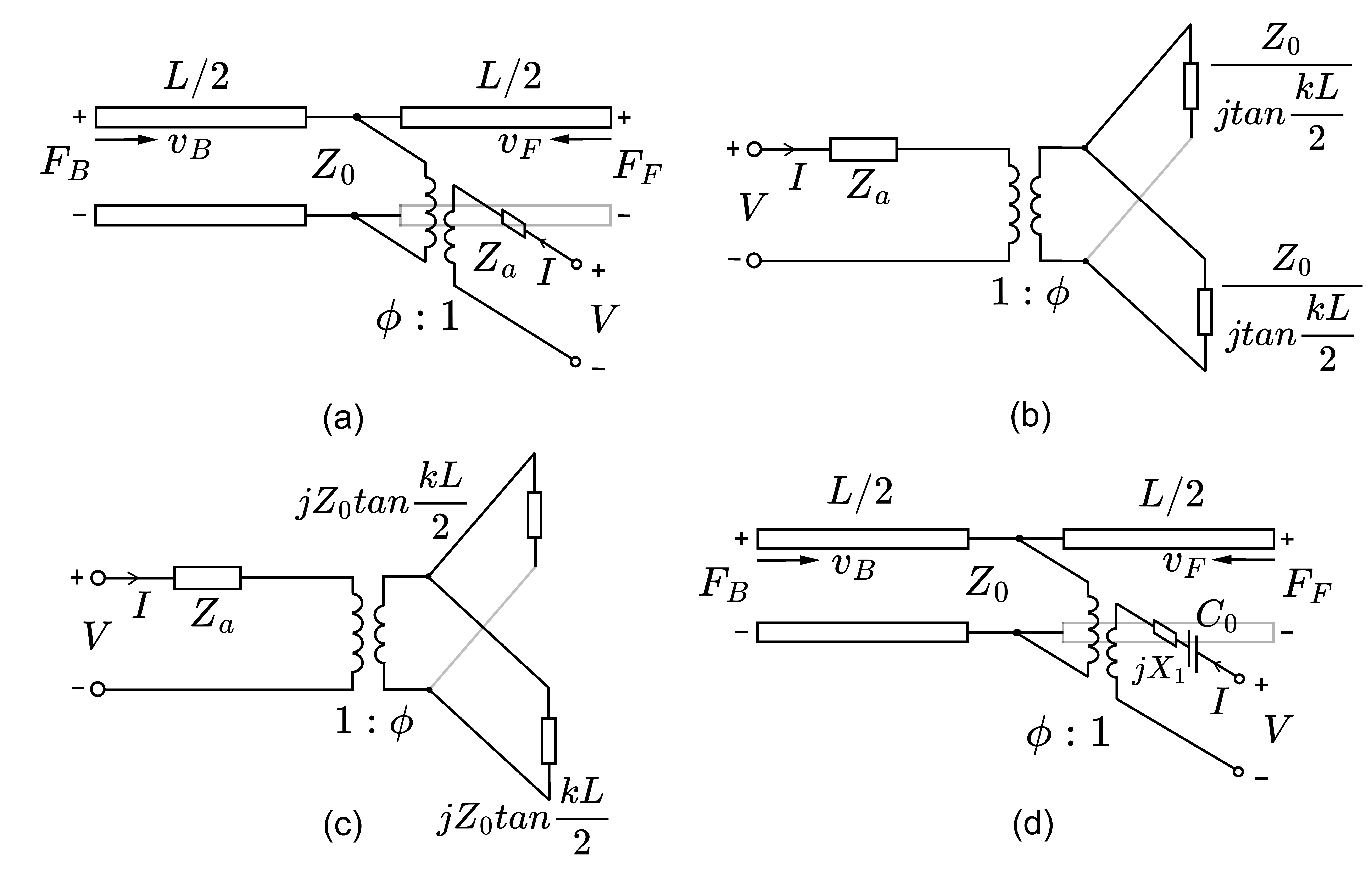}}
\caption{(a) is the general circuit form of the KLM model, (b) and (c) are its simplified forms with open-circuited acoustic ports and short-circuited acoustic ports respectively, (d) is the KLM model with full parameters, where $jX_1= jh_{33}^2\sin{kL}/\omega^2Z_0$.}
\label{Fig2}
\end{figure}

The KLM model exhibits the general form shown in Fig.~\ref{Fig2}(a), where the acoustic transmission line describes the 1D propagation characteristics within the piezoelectric layer; the ideal transformer with turns ratio $\phi$ denotes the conversion between acoustic and electrical quantities; the complex impedance $Z_a$ adds attenuation and phase shift to signals. Thus, the key to deriving the model lies in determining $\phi$ and $Z_a$ to satisfy \eqref{eq1}, \eqref{eq2}, and \eqref{eq3}. This step was accomplished by considering two specific boundary conditions at the acoustic ports.

First, set both acoustic ports to open-circuit conditions, i.e., let $v_B = v_F = 0$ in \eqref{eq3}:
\begin{equation}
    V = \frac{I}{j\omega{C_0}}.\label{eq4}
\end{equation}
Due to the characteristics of open-circuited transmission lines, the circuit in Fig.~\ref{Fig2}(a) can be simplified to the form shown in Fig. ~\ref{Fig2}(b), leading to:
\begin{equation}
    \frac{V}{I} = Z_a + \frac{Z_0}{2\phi^2j\tan{\frac{kL}{2}}}.\label{eq5}
\end{equation}
Combining \eqref{eq4} and \eqref{eq5} yields the following:
\begin{equation}
    Z_a + \frac{Z_0}{2\phi^2j\tan{\frac{kL}{2}}} = \frac{1}{j\omega{C_0}}.\label{eq6}
\end{equation}

Second, set both acoustic ports to a short-circuit state. Substituting $F_B=F_F=0$ into \eqref{eq1}, \eqref{eq2} and \eqref{eq3} yields:
\begin{equation}
    \frac{V}{I} = \frac{2jh_{33}^2\tan{\frac{kL}{2}}}{\omega^2Z_0} + \frac{1}{j\omega{C_0}}.\label{eq7}
\end{equation}
By utilizing the characteristics of the short-circuited transmission lines, Fig.~\ref{Fig2}(a) is simplified to Fig~\ref{Fig2}(c), leading to:
\begin{equation}
    \frac{V}{I} = Z_a + \frac{jZ_0\tan{\frac{kL}{2}}}{2\phi^2}.\label{eq8}
\end{equation}
Combining \eqref{eq7} and \eqref{eq8} yields:
\begin{equation}
    Z_a + \frac{jZ_0\tan{\frac{kL}{2}}}{2\phi^2} = \frac{2jh_{33}^2\tan{\frac{kL}{2}}}{\omega^2Z_0} + \frac{1}{j\omega{C_0}}.\label{eq9}
\end{equation}

Equations \eqref{eq6} and \eqref{eq9} constitute a system of equations with $Z_a$ and $\phi$ as unknowns, solving which completes the derivation (see Fig.~\ref{Fig2}(d)):
\begin{equation}
    \phi=\frac{\omega Z_0}{2h_{33}\sin{\frac{kL}{2}}}.\label{eq10}
\end{equation}
\begin{equation}
    Z_a=\frac{1}{j\omega{C_0}} + \frac{jh_{33}^2\sin{kL}}{\omega^2Z_0}.\label{eq11}
\end{equation}

It is noteworthy that since the system is linear from the piezoelectric constitutive equations to the derived \eqref{eq1}, \eqref{eq2} and \eqref{eq3}, the solutions obtained using open-circuited and short-circuited acoustic ports are universally applicable.

The derivation reveals that the KLM model is fundamentally governed by linear piezoelectric equations and 1D wave equation. Consequently, the physical premises for employing it include the isothermal, adiabatic \cite{b40}, low-frequency \cite{b41}, and small-signal \cite{b42} conditions corresponding to the former, and the TE-mode 1D vibration assumption \cite{b39} corresponding to the latter.

\subsubsection{End-to-end equivalent circuit based on improved KLM model}
The standard KLM model captures the acoustic-electrical conversion but neglects back-end cables and receiver circuits, its acoustic transmission line also lacks intuitiveness when analyzing the effects of critical system-level parameters such as EA, CL, and RI. Leveraging the purely receiving characteristic of PAI, we simplify its acoustic components while extending its electrical components to establish an end-to-end equivalent circuit model.

\begin{figure}[!t]
\centerline{\includegraphics[width=\columnwidth]{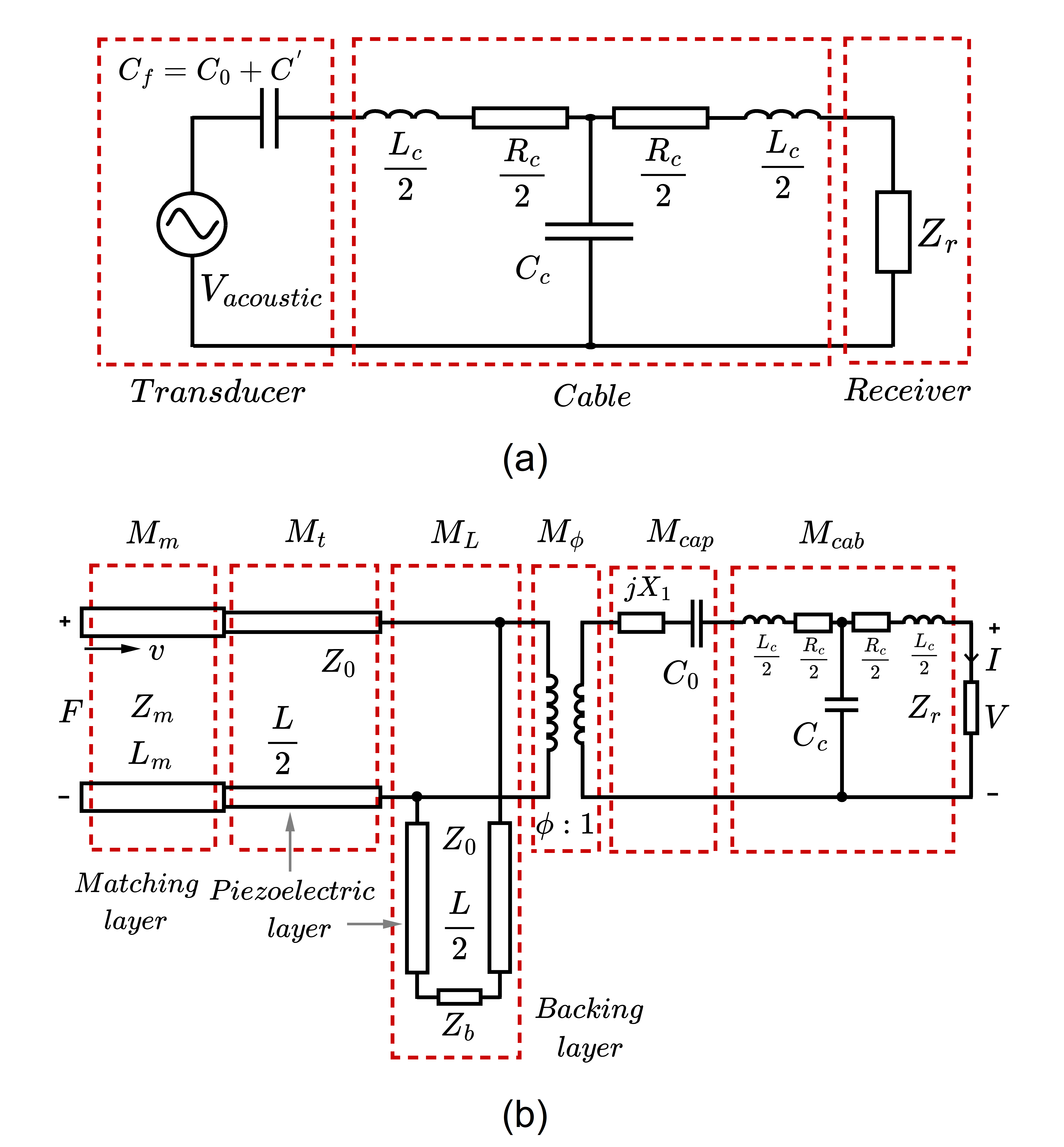}}
\caption{The improved model based on the KLM model. (a) and (b) show the forms of it when calculating detecting sensitivity at a certain frequency and on the entire frequency band respectively.}
\label{Fig3}
\end{figure}

Similar to Thevenin equivalence, we simplify the acoustic components — including the acoustic transmission line and ideal transformer — into a controlled voltage source $V_{acoustic}$ (see Fig.~\ref{Fig3}(a)), whose output corresponds to the open-circuit output voltage in the original KLM model thereby remains independent of the cable and receiver at the load side downstream. In other words, $V_{acoustic}$ depends on the acoustic parameters of each layer in the transducer, as well as the amplitude and frequency of the acoustic excitation, whereas is not directly related to CL or RI. 
\begin{equation}
    c_{33}^DS_3-h_{33}D_3=T_3.\label{eq12}
\end{equation}

On the other hand, \eqref{eq12} shows that the stress $T_3$ — force per unit area — governs the electric displacement $D_3$ and strain $S_3$ in the piezoelectric plate. Therefore, merely increasing EA while maintaining constant surface acoustic pressure does not intensify the electric field, thus cannot raise the open-circuit output voltage. Accordingly, $V_{acoustic}$ is also independent of EA.

So $V_{acoustic}$ can be treated as a constant voltage source when analyzing EA, CL, and RI at a fixed acoustic pressure and frequency.

As with CL and RI, the effect of EA on sensitivity arises mainly through its influence on electrical resonance behavior. Although a larger EA does not strengthen the internal electric field, it increases the plate’s capacitance, enabling more charge storage at the same voltage. This modifies the impedance characteristics of the circuit formed by the piezoelectric capacitance, cable parasitic parameters and receiver, thereby affecting the detected voltage response.

In this study, a lumped T-network representing the cable and a complex receiver impedance were incorporated at the electrical ports (see Fig.~\ref{Fig3}(a)). $C_c$, $R_c$, and $L_c$ denote the cable’s distributed parasitic capacitance, resistance, and inductance, respectively, scaled by CL. This T-network model draws from microwave transmission line theory \cite{b45}, but simplifies it substantially while preserving essential physical effect — such as attenuation, phase shift, and potential resonance — and directly links parasitic parameters to CL. The rationale for treating the cable as a lumped network rests on its physical length being much smaller than the electromagnetic wavelength in the operating band \cite{b46}. In this work, transducers operate at a center frequency of 5 MHz, and cable lengths under 5 m are less than 1/10 of the electromagnetic wavelength ($\approx$~60 m). Given that typical PAI systems operate at frequencies on the order of tens of MHz, this simplification is widely applicable.

Integrating these simplifications and extensions yields a model (see Fig.~\ref{Fig3}(a)) for the end-to-end PAI detection system. Here, $C_0$ is the clamped capacitance of the piezoelectric plate and $\frac{1}{j\omega C^\prime} = \frac{jh_{33}^2\sin{kL}}{\omega^2Z_0}$ represents its capacitance variation due to vibration, corresponding to $jX_1$ in the original KLM model.

\subsubsection{The derivation of transfer function using the improved model}
To further characterize the detection sensitivity using the improved model, this study derives the receiving transfer function of the system. As the output of $V_{acoustic}$ in Fig.~\ref{Fig3}(a) is frequency-dependent, we analyze single-frequency and full-band cases separately to balance simplicity and accuracy.

For a fixed frequency with EA, CL, and RI as independent variables, $V_{acoustic}$ can be regarded as a constant voltage source. Thus, the relative sensitivity is obtained from the transfer function $H_1(\omega)$ between $V_{acoustic}$ and the voltage across the receiver terminals. Based on Fig.~\ref{Fig3}(a), it is given by:
\begin{equation}
    H_1(\omega)=\frac{\frac{1}{j\omega C_c}||Z}{Z-Z_r+\frac{1}{j\omega C_f}+\frac{1}{j\omega C_c}||Z} \cdot\frac{Z_r}{Z}\label{eq13}
\end{equation}
\begin{equation}
    =\frac{Z_r}{\left(Z-Z_r+\frac{1}{j\omega C_f}\right)\left(1+j\omega C_cZ\right)+Z}.\label{eq14}
\end{equation}
Here, $Z=\frac{R_c+j\omega L_c}{2}+Z_r$, and $A||B$ denotes the parallel impedance of A and B. The angular frequency $\omega$ is the target frequency. EA influences the total piezoelectric capacitance $C_f$, while the cable parasitic parameters $C_c$, $R_c$, and $L_c$ are proportional to CL, RI appears explicitly in the expression. The relative sensitivity at a given frequency is obtained by taking the magnitude of Eq.~\eqref{eq14} and normalizing it, allowing systematic optimization of EA, CL, and RI.

In the full-band case, however, $V_{acoustic}$ is not constant, so the sensitivity must be computed using the complete transfer function $H_2(\omega)$ from the acoustic force at the transducer surface to the receiver voltage (see Fig.~\ref{Fig3}(b)). Adopting a transfer matrix approach following~\cite{b21}, we define $M_m$, $M_t$, $M_L$ as acoustic transmission matrices for the matching layer, the adjacent half of the piezoelectric layer, and the parallel combination of the backing layer and the other half of the piezoelectric layer, while $M_\phi$, $M_{cap}$, and $M_{cab}$ are matrices for the ideal transformer, total piezoelectric capacitance (including static and dynamic components), and the lumped cable T-network. The overall transfer matrix from the transducer surface to the receiver is:
\begin{equation}
    M=M_m\cdot M_t\cdot M_L\cdot M_{\phi}\cdot M_{cap}\cdot M_{cab}=
    \begin{pmatrix}
    A & B \\
    C & D
    \end{pmatrix}.\label{eq15}
\end{equation}

Let $F$ and $v$ be the force on the transducer surface and the particle velocity, $V$ and $I$ be the voltage across the receiver impedance and the current through it (complex values, including magnitudes and phases). Then:
\begin{equation}
    F=AV+BI.\label{eq16}
\end{equation}
\begin{equation}
    v=CV+DI.\label{eq17}
\end{equation}
The boundary conditions are:
\begin{equation}
    F=F_i+F_r.\label{eq18}
\end{equation}
\begin{equation}
    v=\frac{1}{Z_c}\left(F_i-F_r\right).\label{eq19}
\end{equation}
\begin{equation}
    I=\frac{V}{Z_r}.\label{eq20}
\end{equation}

Here, $F_i$ and $F_r$ are the complex values denoting incident and reflected force waves, $Z_c$ is the acoustic impedance of the coupling medium, and the sign convention in \eqref{eq19} reflects opposite velocity directions for incident and reflected waves. Eliminating $F_r$ from \eqref{eq18} and \eqref{eq19}, substituting \eqref{eq16}, \eqref{eq19} and \eqref{eq20}, yields the complete transfer function:

\begin{equation}
    H_2(\omega)=\frac{V}{F_i}=\frac{2Z_r}{AZ_r+B+CZ_rZ_c+DZ_c}.\label{eq21}
\end{equation}

The magnitude of \eqref{eq21} gives the full-band detection sensitivity. RI appears explicitly, while EA and CL influence the coefficients $A$, $B$, $C$ and $D$ via $M_{cap}$ and $M_{cab}$, respectively. These can be optimized using the chain rule, though detailed steps are omitted here.

\subsection{Experimental verification}
\subsubsection{Strategy and setup}
To validate the proposed end-to-end model, both electrical–acoustic-source (EAS) and optical–acoustic-source (OAS) experiments were performed.

In EAS experiment, a water-immersed transducer (Olympus V309, 5 MHz) emitted a 10-cycle sinusoidal pulse train at 5 MHz. This setup provided consistent single-frequency performance and enabled clear separation of electrical crosstalk from the target signal based on time-of-arrival. Although limited in bandwidth, EAS ensured high pulse reproducibility across batches, making it suitable for comparative analysis of relative sensitivity under different system parameters at a fixed frequency.

OAS experiment utilized 10-ns laser pulses to excite a black-coated quartz block, generating broadband acoustic signals with a flat spectral response across typical PAI frequencies. This approach allowed measurement of the amplitude–frequency response and bandwidth under varying parameters. However, temporal fluctuations in laser pulse energy and spot size — due to thermal instability — precluded direct comparison of raw signal amplitudes acquired at different times.

The tested probes were circular piezoelectric composite transducers (ULSO, 5 MHz) with diameters of 1.5, 2, 2.5, 3, and 4 mm. All parameters other than diameter were held constant, and each probe had a 1.5 m uniform cable. By adding coaxial cables of the same model and SMA connectors, five CL values were obtained: 1.5, 2, 2.5 3, and 3.5 m. A custom receiver circuit was designed, with the ground resistance of the first-stage amplifier adjusted to vary the input impedance. Using a vector network analyzer (VNA, Keysight E5071C), the input impedances of the four receiver channels at 5 MHz were measured as:
$404-j\ast324~\Omega$, $145-j\ast24~\Omega$, $128-j\ast17~\Omega$, and $51-j\ast0.07~\Omega$ for channels 1, 2, 3, and 4 respectively. Parasitic capacitance, present in all channels, is not a design flaw but a common non-ideal characteristic arising from physical circuit structures such as bond wires, solder balls, and coupled metal layers in the package substrate—an inevitable result of circuit implementation \cite{b46}.

Parameter calibration was performed using multiple methods to ensure simulation accuracy. $Z_r$ was set to the VNA-measured values. Distributed cable capacitance, resistance, and inductance, as well as piezoelectric element capacitance, were cross-verified by at least two of the following: VNA measurement, multimeter measurement (VICTOR VC890F), or calculation based on manufacturer specifications. Material parameters for matching, piezoelectric, and backing layers were provided by the manufacturer.

Given the complex coupling among EA, CL, and RI, each parameter was controlled individually in EAS experiments to analyze interactions between the other two. Initially, EA–RI coupling was examined to optimize the acoustic–electrical conversion efficiency. Subsequently, CL–RI coupling was studied to minimize losses in the electrical signal transmission path.

\subsubsection{Results and analysis}
\label{sec:core conlusion}
\begin{figure}[!t]
\centerline{\includegraphics[width=\columnwidth]{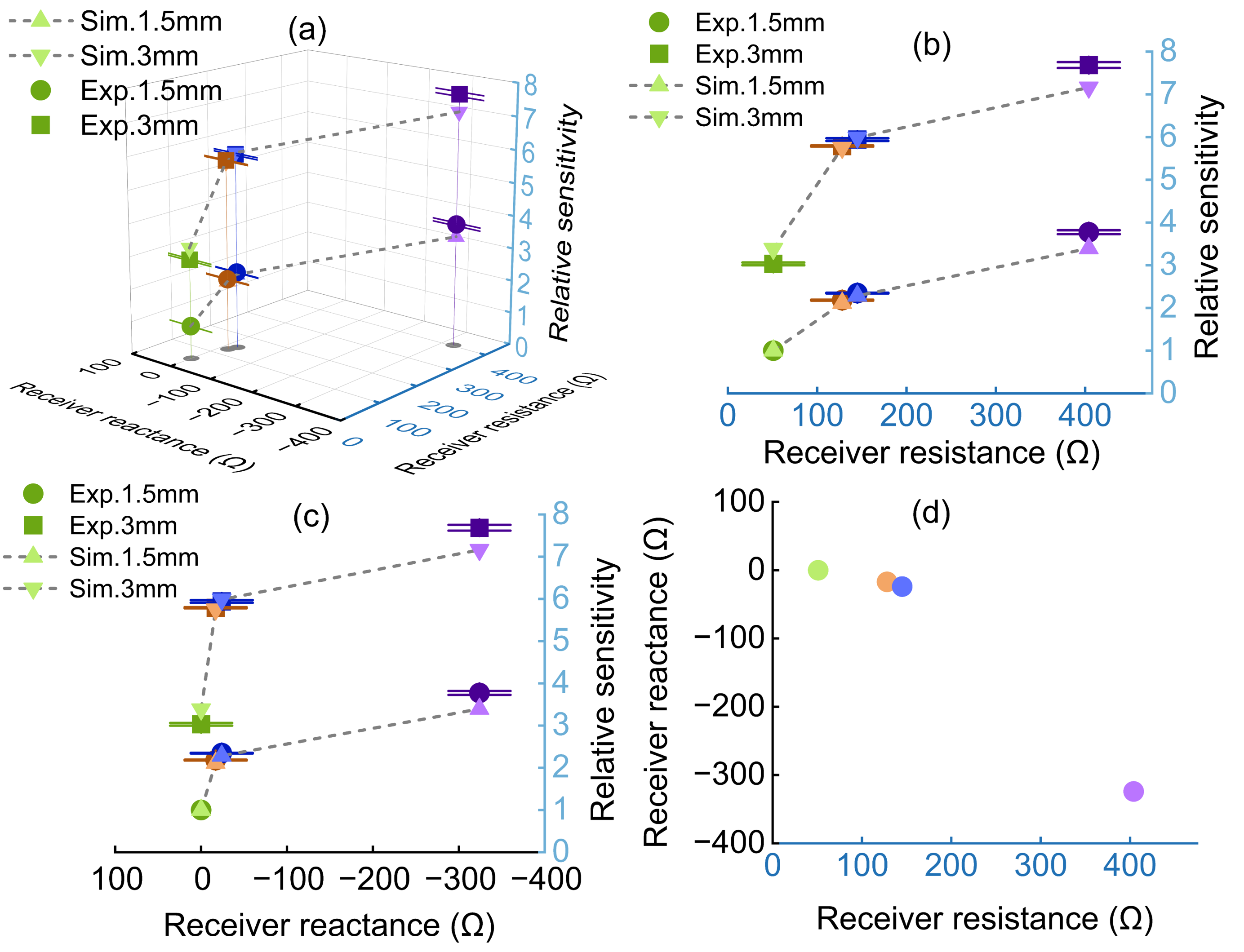}}
\caption{The influence of EA and RI on detection sensitivity and their coupling effects in both simulation and experiment. (a) is the 3D-plot in which X, Y, Z axis represent resistance, reactance of the receiver and relative sensitivity respectively, (b), (c), and (d) are the projections of (a) onto the three planes. purple, blue, orange and green denote channel 1, 2, 3, and 4.}
\label{Fig4}
\end{figure}

To demonstrate the coupling among EA, RI and their influence on detection sensitivity, we fixed CL at 1.5 m and employed two test transducers with diameters of 1.5 mm and 3 mm, corresponding to an EA ratio of 1:4. These were connected to four receiver channels for EAS experiments at eight parameter points. Simulation data were derived from \eqref{eq13}. Received voltage amplitudes from both experiments and simulations were normalized to the value obtained with the 1.5 mm transducer connected to channel 4. Experimental results agreed well with simulations (average relative error $<5\%$), confirming the validity of the improved model.

With RI and CL fixed, increasing EA significantly improves detection sensitivity (Fig.~\ref{Fig4}). However, the extent of this improvement depends on the other parameters. For instance, quadrupling EA enhanced sensitivity by factors of approximately 2, 1.7, 1.5, and 1 for channels 4 to 1, respectively. The effect is more pronounced under low RI conditions. According to the improved model, low RI results in a lower total parallel impedance with the cable T-network, making voltage division more sensitive to variations in piezoelectric capacitance. Similarly, when cable parasitic capacitance is comparable to piezoelectric capacitance, EA exerts a greater influence on sensitivity.

With EA and CL fixed, increasing RI also enhances sensitivity (see Fig.~\ref{Fig4}), though with diminishing returns. For the 1.5 mm probe, sensitivity increased by about 1.3 times when RI rose from channel 4 ($|Z_r|\approx51~\Omega$) to channel 2 ($|Z_r|\approx147~\Omega$), but only 0.5 times from channel 2 to channel 1 ($|Z_r|\approx518~\Omega$). A similar trend occurred with the 3 mm probe. The improved model explains that at high RI, the system approaches an open-circuit condition, with sensitivity converging to the theoretical limit $|V_{acoustic}|$, thus reducing the benefit of further RI increases.

This analysis informs the design of spatially constrained systems. When EA must be minimized due to invasiveness or portability, RI can be raised to offset sensitivity loss. However, excessively high RI yields marginal gains and may introduce artifacts (see Section~\ref{sec:radial-resonance}). System design should thus balance clinical requirements.

To examine CL–RI coupling, a 2 mm probe was tested at five CL values (1.5–3.5 m), each connected to four channels, yielding 20 EAS measurements. Simulated data from \eqref{eq13} were normalized to the CL = 3.5 m case. Experiments closely matched simulations (mean relative errors for channels 1–4: $1.30\%$, $4.97\%$, $2.47\%$, and $1.39\%$, respectively, all $<5\%$), further validating the model.

\begin{figure}[!t]
\centerline{\includegraphics[width=\columnwidth]{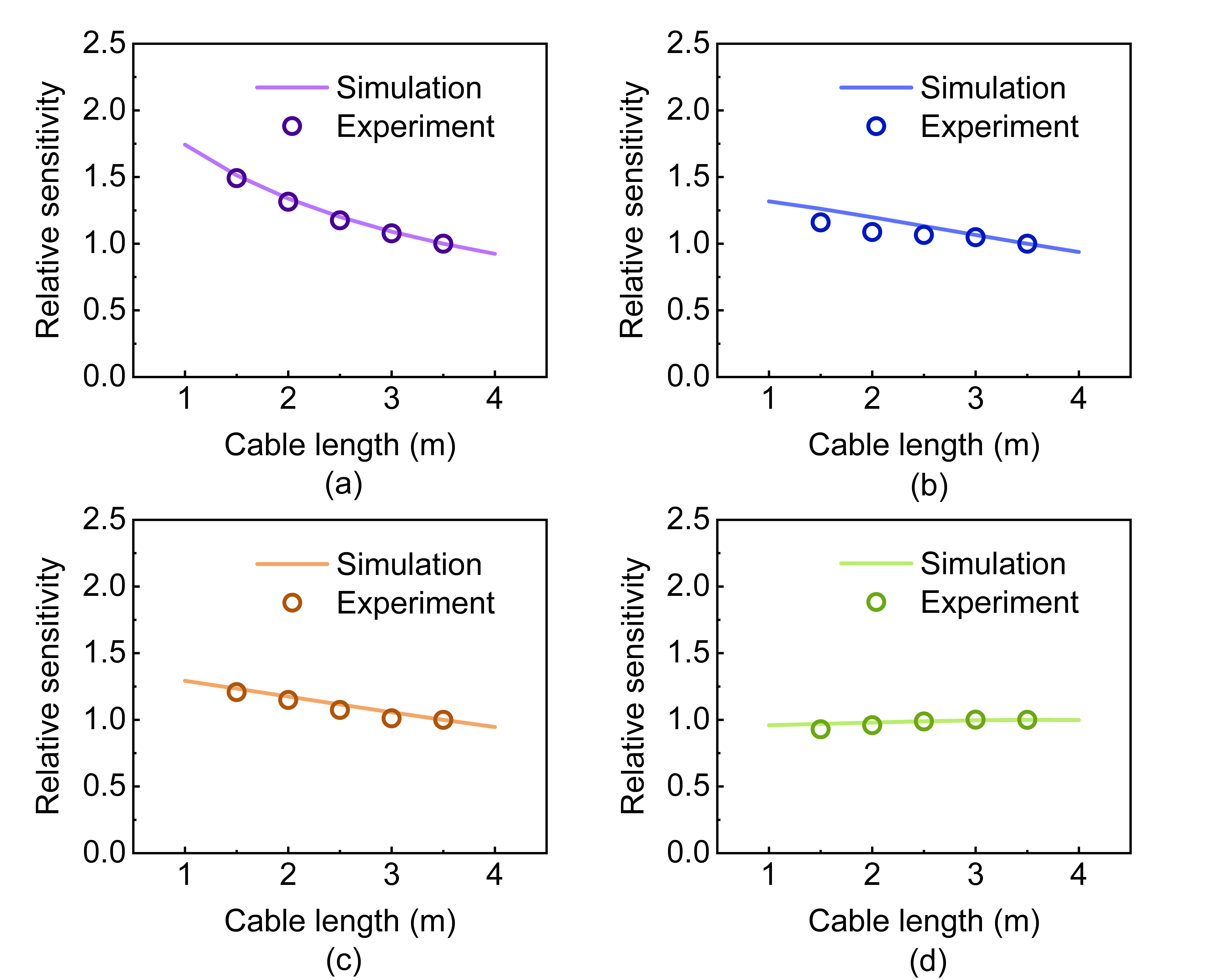}}
\caption{The influence of CL and RI on detection sensitivity and their coupling effects in both simulation and experiment. (a), (b), (c), and (d) denote channel 1, 2, 3, and 4 respectively, corresponding to purple, blue, orange, and green.}
\label{Fig5}
\end{figure}

When EA is fixed and RI is high, detection sensitivity decreases monotonically with CL (see Fig.~\ref{Fig5}), and higher RI increases the sensitivity of it to CL variations: increasing CL from 1.5 m to 3.5 m reduced sensitivity by about $34\%$, $21\%$, and $19\%$ for channels 1–3, respectively. Near-linear trends in channels 2 and 3 result from their small rate of change. The improved model explains that higher RI makes total impedance more sensitive to parasitic capacitance changes in the cable T-network, amplifying the effect of CL on sensitivity.

Under fixed EA and low RI, sensitivity varies non-monotonically with CL (see Fig.~\ref{Fig5}). For channel 4, sensitivity peaks between 3 m and 3.5 m. The model indicates that appropriate CL adjusts the capacitive–inductive reactance ratio, moving the circuit closer to resonance. Despite this non-monotonic behavior, sensitivity changes more smoothly at low RI: channel 4 increased by only about $5\%$ over the CL range, much less than channels 1–3. The model's accurate fit for minute non-monotonic variation confirms that the lumped T-network effectively represents cable behavior in PAI detection systems.

Thus, for high RI ($|Z_r|>100~\Omega$), CL should be minimized to avoid sensitivity loss. For low RI ($|Z_r|<100~\Omega$), the optimal CL can be determined via simulation by our improved model.

To assess RI's effect on sensitivity spectrum and bandwidth, a 3 mm probe was used in OAS experiments with CL = 1.5 m. IR was measured across channels 1--4, and waveforms (40 MHz sampling) were processed via Discrete Fourier Transform (DFT) to obtain the amplitude--frequency response from 0--20 MHz. Simulated spectra from \eqref{eq21} closely matched measurements in the 2--20 MHz band, with mean absolute errors of $3.6\%$--$4.8\%$ of each channel’s sensitivity peak ($<5\%$). Notably, magnitude approached or equaled zero at certain high-frequency points, so relative error calculations were meaningless.

\begin{figure}[!t]
\centerline{\includegraphics[width=\columnwidth]{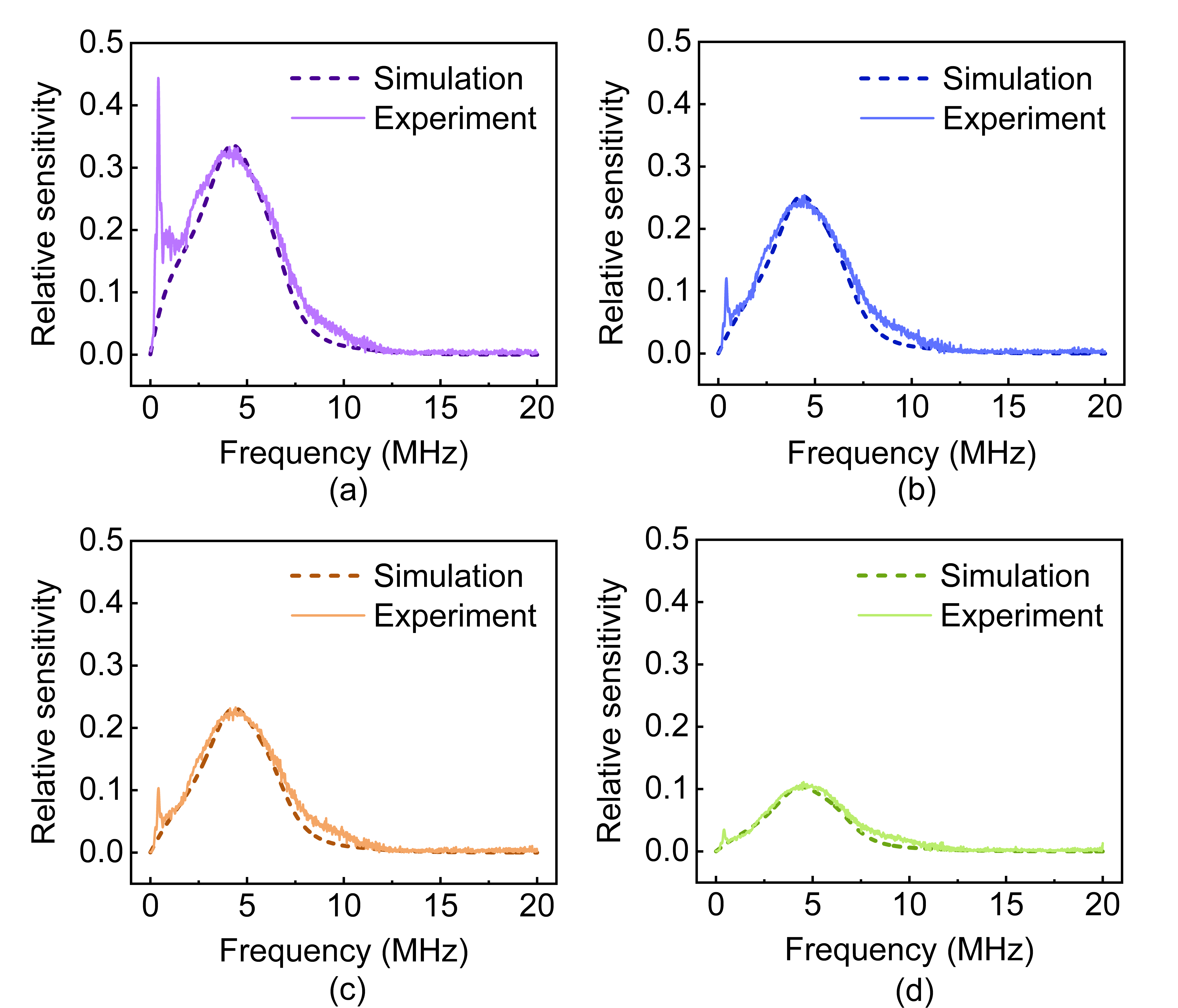}}
\caption{The impact of RI on the amplitude-frequency response and bandwidth of the detection system in both simulation and experiment. (a), (b), (c), and (d) denote channel 1, 2, 3, and 4 respectively, corresponding to purple, blue, orange, and green.}
\label{Fig6}
\end{figure}

Increasing RI significantly enhances low-frequency response while having a relatively minor impact on high-frequency response (see Fig.~\ref{Fig6}). For example, raising RI from channel 4 to channel 1 improved sensitivity by about 3.1 times at 2.5 MHz versus about 2.1 times at 7.5 MHz.

Higher RI also broadens system bandwidth: the –6 dB bandwidth increased from 4.27 MHz to 4.82 MHz ($\approx13\%$) from channel 4 to channel 1. The aforementioned patterns of high-frequency and low-frequency response variations with RI changes is the main reason for this bandwidth expansion, the –6 dB center frequency shifted from 4.60 MHz to 4.19 MHz, confirming this observation. The model predicts that at $k\Omega$-level RI, an additional left-side peak emerges, further extending bandwidth.

In summary, high RI improves low-frequency response and bandwidth, beneficial for deep-tissue imaging such as breast or transcranial PAI, where low-frequency acoustic signals suffer less attenuation. High RI also benefits multispectral PAI by maximizing bandwidth for heterogeneous signal acquisition.

\subsection{Noise Analysis}

Under typical operating conditions of PAI acoustic detection systems, 1/f noise and shot noise are negligible. Thus, Johnson thermal noise from both the receiver and the transducer (collectively termed thermal noise) represents the dominant system noise source \cite{b20}. The total noise level is primarily influenced by EA and RI. EA modifies the electrical impedance of the transducer, directly contributing to its thermal noise and indirectly influencing receiver thermal noise through electrical coupling with back-end circuitry. RI, in turn, directly affects the receiver's thermal noise.

\begin{table}
\caption{Noise PSD}
\setlength{\tabcolsep}{3pt}
\begin{tabular}{
p{74pt}
p{34pt}
p{34pt}
p{34pt}
p{34pt}
}
\hline\hline
Experiment \par ($\times10^{-18}V^2$/Hz) & channel 4 & channel 3 & channel 2 & channel 1 \\
\hline
3mm-diameter \par probe & 4.03 & 6.08 & 5.30 & 6.49 \\
\hline
1.5mm-diameter \par probe & 4.40 & 4.28 & 4.93 & 6.23 \\
\hline\hline
Simulation \par ($\times10^{-18}V^2$/Hz) & channel 4 & channel 3 & channel 2 & channel 1 \\
\hline
3mm-diameter \par probe & 1.56 & 1.95 & 2.01 & 2.49 \\
\hline
1.5mm-diameter \par probe & 1.41 & 1.60 & 1.63 & 1.78 \\
\hline\hline
\end{tabular}
\label{tab1}
\end{table}

Table~\ref{tab1} summarizes experimental and simulated values of the average noise power spectral density (PSD) over 0--20 MHz, corresponding to the eight parameter points in Fig.~\ref{Fig4} (two EA values, four RI values, CL fixed at 1.5 m). Measurements were performed in a microwave anechoic chamber with idle probes - not immersed in water tank or excited - connected to the receiver via their respective cables. Noise voltage signals were recorded at the receiver, and their PSD was computed and averaged across 0--20 MHz. Simulations were based on the noise model of the receiver amplifier (LMH6629, Texas Instruments) in its datasheet, with transducer impedance estimated using our improved model.

Experimental and simulated noise PSD values show generally consistent trends and are of the same order of magnitude, though measured values exceed simulations at all parameter points. Possible reasons include: additional noise from PCB layout inaccuracies in the physical circuit; incomplete modeling of reactance effects on the thermal noise spectrum; power supply noise from insufficient filtering; and common-mode noise due to non-ideal grounding.

\begin{figure}[!t]
\centerline{\includegraphics[width=\columnwidth]{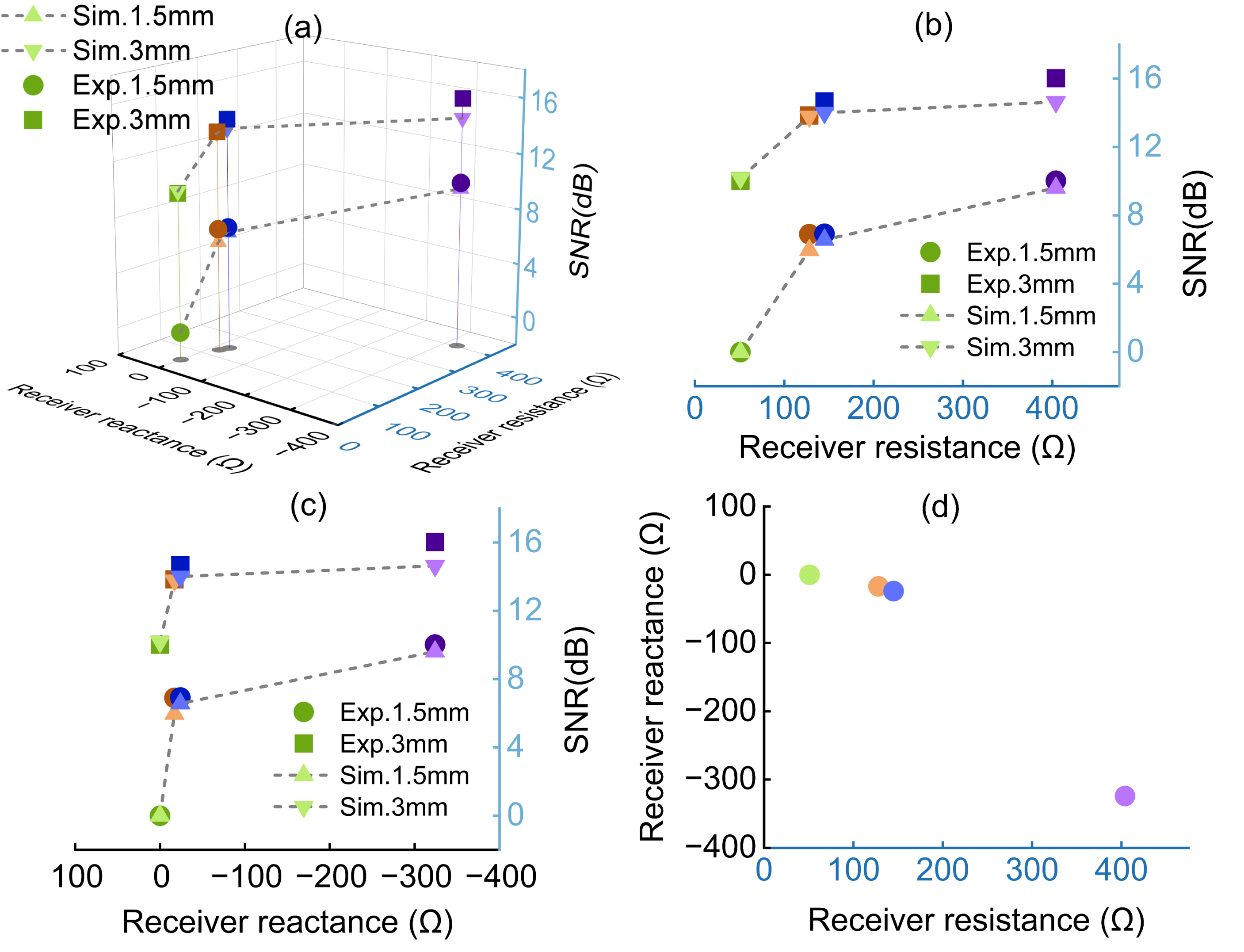}}
\caption{The influence of EA and RI on SNR in both simulation and experiment. (a) is the 3D-plot in which X, Y, Z axis represent resistance, reactance of the receiver and SNR respectively, (b), (c), and (d) are the projections of (a) onto the three planes. Purple, blue, orange, and green denote channel 1, 2, 3, and 4.}
\label{Fig7}
\end{figure}

To examine how experimental and simulated signal-to-noise ratio (SNR) vary with EA and RI, noise data are incorporated into the sensitivity results from Fig.~\ref{Fig4}. The normalized sensitivity values serve as the signal component. Noise amplitudes are derived as the square root of the values in Table~\ref{tab1}, then normalized using the result from the 1.5 mm probe in channel 4 (this shifts experimental and simulated SNR curves closer without altering their shapes, aiding trend comparison). As shown in Fig.~\ref{Fig7}, increasing EA or RI improves not only detection sensitivity but also SNR. At fixed EA and CL, SNR rises with RI, though the rate of increase gradually diminishes. Similarly, at fixed RI and CL, SNR increases with EA, with a more pronounced effect under low impedance conditions.

\section{Further discussion on boundaries of model applicability}
\label{sec:radial-resonance}
Our improved model inherits the limitations of the underlying KLM model, including isothermal, adiabatic \cite{b40}, low-frequency \cite{b41}, and small-signal \cite{b42} conditions, as well as 1D vibration assumption \cite{b39}. Real-world scenarios often exceed these idealizations, such as nonlinear effects under high-frequency excitation, multi-dimensional vibration coupling, and saturation under large-signal drive. A representative example is the low-frequency tailing phenomenon, which arises from the breakdown of the 1D vibration assumption. The following section analyzes its origin to underscore the importance of recognizing model limitations.

\subsection{Manifestations and impacts of the tailing phenomenon}
\begin{figure}[!t]
\centerline{\includegraphics[width=\columnwidth]{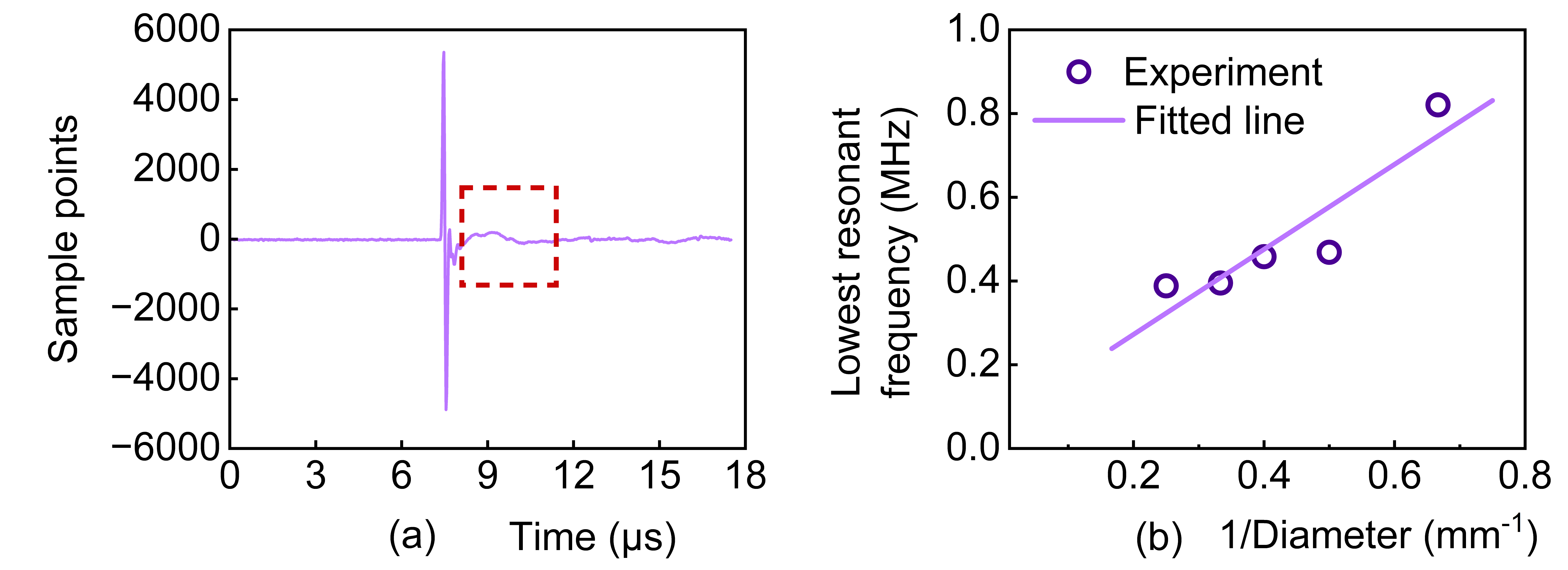}}
\caption{(a) shows the low-frequency tailing in IR waveform, (b) illustrates the relationship between lowest resonant frequency and the diameter of the probes.}
\label{Fig8}
\end{figure}

All experimentally measured spectra exhibit a narrow low-frequency peak not captured by the improved model (see Fig.~\ref{Fig6}). This peak increases in amplitude with rising RI, whereas its center frequency remains near 0.4 MHz. In the time domain, it appears as a long tail following the main IR signal (see Fig.~\ref{Fig8}(a)).

This tailing phenomenon degrades PA image quality by introducing low-frequency block artifacts around highly absorbing structures such as blood vessels and organs. These artifacts obscure anatomical boundaries and generate spurious signals in otherwise homogeneous regions, potentially leading to misdiagnosis of diffuse pathologies. Identifying the origin of this tailing is therefore essential.

\begin{figure*}[!t]
\centerline{\includegraphics[width=\textwidth]{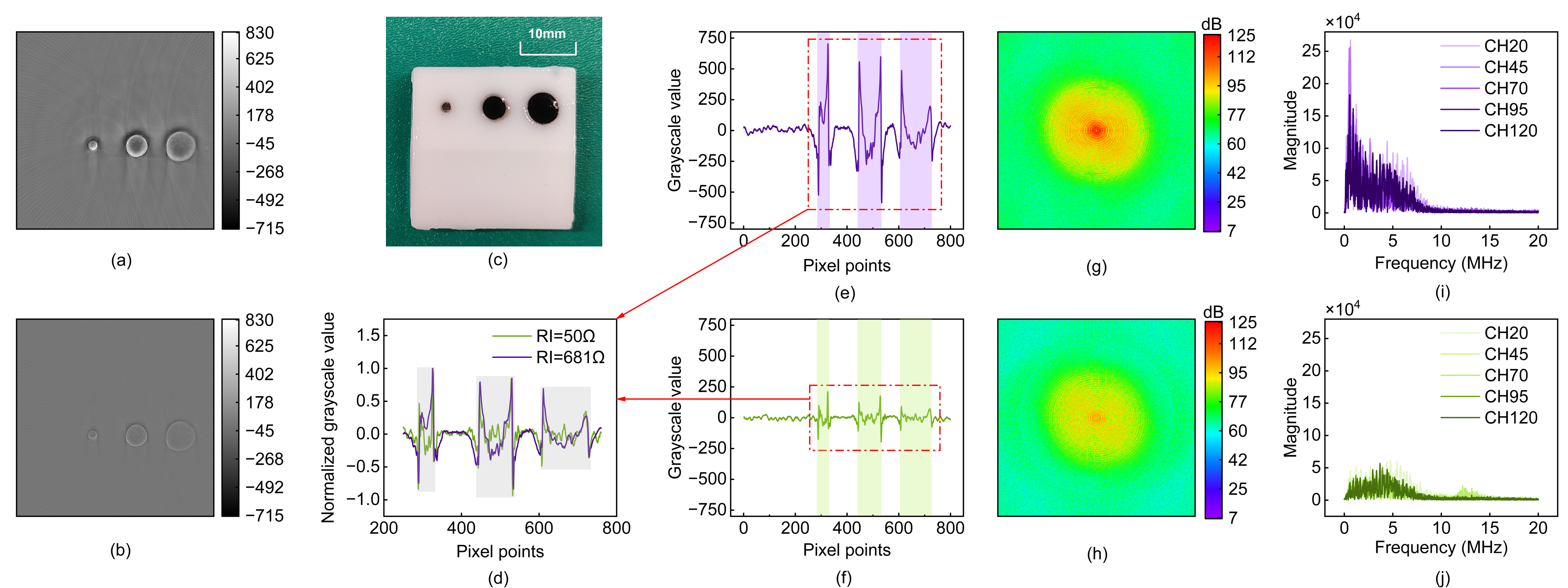}}
\caption{(a) and (b) show the DAS-reconstructed PA image with RI=681 $\Omega$ and 50 $\Omega$ respectively, (c) is a photo of the phantom, (e) and (f) display the grayscale value profiles along the diametrical cross-sections passing through the three circles in (a) and (b), respectively; (d) presents a comparison of signal patterns by showing the normalized grayscale values from (e) and (f); (g) and (h) show the results of the 2D-DFT applied to (a) and (b), respectively; finally, (i) and (j) present the magnitude spectra of the raw signals acquired at channels 20, 45, 70, 95, and 120 from the sinograms (acquired with 128 channels in total) corresponding to (a) and (b), respectively.}
\label{Fig9}
\end{figure*}

\subsection{Source of the tailing phenomenon}
\subsubsection{Radial resonance mode hypothesis}
One possible cause is Gibbs oscillation due to finite system bandwidth. However, such oscillations typically occur near the transducer's center frequency (5 MHz in this study), whereas the observed peak lies at around 0.4 MHz, ruling out this explanation.

We propose that the tailing stems from standing wave modes associated with radial vibration mode of the piezoelectric plate. The specific mechanism is as follows: when a longitudinal plane wave excites the plate vertically, thickness-mode vibrations propagate radially as shear waves. Additionally, acoustic impedance mismatch at the radial boundaries — such as between the piezoelectric element and the stainless steel housing in single-element transducers, or within unfilled cutting gaps in array transducers — creates reflection conditions conducive to standing wave formation.

These resonances behaviors occur at eigenfrequencies given by: 
\begin{equation}
    f_n=\frac{v_s j_{0,n}}{\pi d}.\label{eq22}
\end{equation}
Where $v_s$, $d$ and $j_{0,n}$ denote shear wave velocity, diameter of the piezoelectric plate, and the $n$th root of the zero-order Bessel function respectively. We define these standing wave modes as radial resonances, such resonances enhance the piezoelectric response at matching frequencies, producing spectral peaks and corresponding time-domain tailing.

\subsubsection{Experimental verification}
Since the manufacturer did not supply the shear wave velocity $v_s$ and given dimensional tolerances in piezoelectric plate diameters $d$, we did not directly compare measured peak frequencies with results from \eqref{eq22}. Instead, after an initial estimate using \eqref{eq22}, we quantitatively examined the relationship between resonant frequency and plate diameter.

For the 3 mm-diameter probe, the measured resonance is around 0.4 MHz. Substituting it into \eqref{eq22} with $n=1$ yields $v_s\approx$ 1568 m/s, consistent with typical values for PZT/epoxy 1-3 composites. 

We then performed OAS experiments using aforementioned circular probes of diameters 1.5--4 mm, all connected to channel 1 (best low-frequency response) with CL fixed at 1.5 m. Notably, the 4 mm-diameter probe exhibited multiple resonant peaks in the low-frequency range, supporting the radial resonance hypothesis: its lower fundamental frequency $f_1$ allows higher-order modes to become distinctly observable. To ensure consistent comparison of the fundamental mode across probes, the lowest resonant frequency was identified in each case, with the center frequency determined using -6 dB center frequency.

\subsubsection{Results}
A linear fit of the lowest resonant frequency against the inverse plate diameter (see Fig.~\ref{Fig8}(b)) yields a slope of 1.016, intercept of 0.069, and coefficient of determination $R^2\approx0.83$, confirming a near-inverse proportionality. Minor deviations are attributable to manufacturing tolerances of the plates and spectral estimation errors, and remain within acceptable limits. These results validate the relationship in \eqref{eq22} and support our radial resonance hypothesis.

\section{Practical imaging verification}
\subsection{Strategy and setup}

To experimentally validate the impact of RI on the sensitivity and frequency response of the detection system, a phantom experiment was conducted. The phantom consisted primarily of white agar (doped with 1/30 volume fraction of intralipid), embedded with black agar cylinders (doped with 1/4 volume fraction of carbon ink) of diameters 2 mm, 4 mm, and 6 mm (see Fig.~\ref{Fig9}(c)). This design achieved distinct optical absorption contrasts whereas maintaining similar acoustic properties between the two materials.

The phantom was irradiated by a laser (Beamtech, Mianna-Q) with wavelength of 1064 nm, pulse duration of 10 ns and repetition rate of 10 Hz. The laser beam was coupled into a light guide, expanded by a concave lens, and vertically illuminated the phantom surface to ensure uniform excitation. PA signals were detected using a semi-ring array transducer (ULSO, 128 elements, 5.5 MHz). RI was varied by connecting the transducer sequentially to two receivers (Marsonics, 80 MHz sampling rate) with input impedance of 50 $\Omega$ and 681 $\Omega$, respectively.

\subsection{Results and analysis}
Delay-and-sum (DAS) reconstruction was performed on data acquired under both RI conditions (see Fig.~\ref{Fig9}(a) and (b)). Evidently, the PA image obtained with higher RI exhibited significantly improved contrast and clearer features, attributable to the enhanced detecting sensitivity under high RI (see Section~\ref{sec:core conlusion}). This was corroborated by the signal amplitude profiles across the diameters of the three cylinders (see Fig~\ref{Fig9}(e) and (f)), where the peak amplitude at 681 $\Omega$ was approximately three times that at 50 $\Omega$.

Moreover, under high RI, the 2 mm-diameter target appeared brighter than the background, whereas the central regions of the 4 mm and 6 mm targets appeared darker. In contrast, under low RI, all three targets exhibited minimal contrast against the background. Additionally, a larger area of magnitude fluctuation around the target boundaries was observed under high RI, resulting in visually thicker contours. These observations are further clearly illustrated by the normalized magnitude profiles across the targets (see Fig.~\ref{Fig9}(d)). Thus, high RI not only improved image contrast via enhanced sensitivity but also induced a pattern change in the image domain.

We attribute these effects primarily to the significant enhancement of the low-frequency response and the increased bandwidth of the detection system under high RI (see Section~\ref{sec:core conlusion}). To verify this hypothesis, 2D-DFT was applied to the PA images (see Fig.~\ref{Fig9}(g) and (h)), and DFT was performed on raw signals from five representative transducer channels under both RI conditions (see Fig.~\ref{Fig9}(i) and (j)). The results clearly indicate that high RI substantially enhanced the low-frequency components in both the PA images and the raw signals, whereas causing relatively minor changes in the high-frequency components.

\section*{Acknowledgment}
The authors would like to thank Guangyu Zeng from Weiyang College, Tsinghua University, for his support in theoretical derivation of radial resonances hypothesis.


\begin{thebibliography}{00}

\bibitem{b1} J. Yao and L. V. Wang, ``Sensitivity of photoacoustic microscopy,'' \emph{Photoacoustics}, vol. 2, no. 2, pp. 87--101, Apr. 2014, doi: 10.1016/j.pacs.2014.04.002.

\bibitem{b2} X. Zhong et al., ``Free-moving-state microscopic imaging of cerebral oxygenation and hemodynamics with a photoacoustic fiberscope,'' \emph{Light Sci. Appl.}, vol. 13, Jan. 2024, Art. no. 5, doi: 10.1038/s41377-023-01348-3.

\bibitem{b3} Y. Liang et al., ``Optical-resolution functional gastrointestinal photoacoustic endoscopy based on optical heterodyne detection of ultrasound,'' \emph{Nat. Commun.}, vol. 13, Dec. 2022, Art. no. 7604, doi: 10.1038/s41467-022-35259-5.

\bibitem{b4} Y. Hazan, A. Levi, M. Nagli, and A. Rosenthal, ``Silicon-photonics acoustic detector for optoacoustic micro-tomography,'' \emph{Nat. Commun.}, vol. 13, Mar. 2022, Art. no. 1488, doi: 10.1038/s41467-022-29179-7.

\bibitem{b5} R. Ansari, E. Z. Zhang, A. E. Desjardins, and P. C. Beard, ``Alloptical forward-viewing photoacoustic probe for high-resolution 3D endoscopy,'' \emph{Light Sci. Appl.}, vol. 7, Oct. 2018, Art. no. 75, doi: 10.1038/s41377-018-0070-5.

\bibitem{b6} X. Tang et al., ``High sensitivity photoacoustic imaging by learning from noisy data,'' \emph{IEEE Trans. Med. Imaging}, vol. 44, no. 7, pp. 2868–-2877, Jul. 2025, doi: 10.1109/TMI.2025.3552692.

\bibitem{b7} S. Hu, K. Maslov, and L. V. Wang, ``Second-generation optical-resolution photoacoustic microscopy with improved sensitivity and speed,'' \emph{Opt. Lett.}, vol. 36, no. 7, pp. 1134–-1136, Apr. 2011, doi: 10.1364/OL.36.001134.

\bibitem{b8} J. Yao and L. V. Wang, ``Photoacoustic microscopy,'' \emph{Laser Photon. Rev.}, vol. 7, no. 5, pp. 758–-778, Sep. 2013, doi: 10.1002/lpor.201200060.

\bibitem{b9} D. Razansky, J. Baeten, and V. Ntziachristos, ``Sensitivity of molecular target detection by multispectral optoacoustic tomography (MSOT),'' \emph{Med. Phys.}, vol. 36, no. 3, pp. 939–-945, Mar. 2009, doi: 10.1118/1.3077120.

\bibitem{b10} A. M. Winkler, K. Maslov, and L. V. Wang, ``Noise-equivalent sensitivity of photoacoustics,'' \emph{J. Biomed. Opt.}, vol. 18, no. 9, Sep. 2013, Art. no. 097003, doi: 10.1117/1.JBO.18.9.097003.

\bibitem{b11} C. Yang et al., ``Sensitivity enhanced photoacoustic imaging using a high-frequency PZT transducer with an integrated front-end amplifier,'' \emph{Sensors}, vol. 20, no. 3, Jan. 2020, Art. no. 766, doi: 10.3390/s20030766.

\bibitem{b12} B. Fu et al., ``Optical ultrasound sensors for photoacoustic imaging: a narrative review,'' \emph{Quant. Imaging Med. Surg.}, vol. 12, no. 2, pp. 1608--1631, Feb. 2022, doi: 10.21037/qims-21-605.

\bibitem{b13} R. Manwar, K. Kratkiewicz, and K Avanaki, ``Overview of ultrasound detection
technologies for photoacoustic imaging,'' \emph{Micromachines}, vol. 11, no. 7, Jul. 2020, Art. no. 692, doi: 10.3390/mi11070692.

\bibitem{b14} A. A. Oraevsky and A. A. Karabutov, ``Ultimate sensitivity of time-resolved optoacoustic detection,'' in \emph{Proc. SPIE}, vol. 3916, \emph{Biomedical Optoacoustics}, San Jose, CA, USA, 2000, doi: 10.1117/12.386326.

\bibitem{b15} R. Krimholtz, D.A. Leedom, and G. L. Matthaei, ``New equivalent circuit for elementary piezoelectric transducers,'' \emph{Electron. Lett.}, vol. 6, no. 13, pp. 398--399, Jun. 1970, doi: 10.1049/el:19700280.

\bibitem{b16} Biosono Technology Co., Ltd., \emph{Transducer KLM Model Simulation}. \text{[Online]}. Available: \url{https://testbuild.biosono.com/?page_id=105}

\bibitem{b17} E.J.W. Merks, J.M.G. Borsboom, N. Bom, A.F.W. van der Steen, and N. de Jong, ``A KLM–circuit model of a multi-layer transducer for acoustic bladder volume measurements,'' \emph{Ultrasonics}, vol. 44, Supplement, pp. e705--e710, Dec. 2006.

\bibitem{b18} G. R. Lockwood and F. S. Foster, ``A coupled two-port network transducer model,'' in \emph{Proc. IEEE Int. Ultrason. Symp. (IUS)}, vol. 1, Tucson, AZ, USA, 1992, pp. 585--588, doi: 10.1109/ULTSYM.1992.275933.

\bibitem{b19} G. Kim, M. K. Seo, N. Choi, K. S. Baek, and K. B. Kim, ``Application of KLM model for an ultrasonic through-transmission method,'' \emph{Int. J. Precis. Eng. Manuf.}, vol. 20, pp. 383–-393, Feb. 2019, doi: 10.1007/s12541-019-00050-y.

\bibitem{b20} C. G. Oakley, ``Calculation of ultrasonic transducer signal-to-noise ratios using the KLM model,'' \emph{IEEE Trans. Ultrason., Ferroelectr., Freq. Control}, vol. 44, no. 5, pp. 1018--1026, Sep. 1997, doi: 10.1109/58.655627.

\bibitem{b21} M. Castillo, P. Acevedo, and E. Moreno, ``KLM model for lossy piezoelectric transducers,'' \emph{Ultrasonics}, vol. 41, no. 8, pp. 671--679, Nov. 2003, doi: 10.1016/S0041-624X(03)00101-X.

\bibitem{b22} J. K. Woodacre, T. G. Landry, and J. A. Brown, ``A low-cost miniature histotripsy transducer for precision tissue ablation,'' \emph{IEEE Trans. Ultrason., Ferroelectr., Freq. Control}, vol. 65, no. 11, pp. 2131--2140, Nov. 2018, doi: 10.1109/TUFFC.2018.2869689.

\bibitem{b23} G. E. Stocker, M. Zhang, Z. Xu, and T. L. Hall, ``Endocavity histotripsy for efficient tissue ablation–transducer design and characterization,'' \emph{IEEE Trans. Ultrason., Ferroelectr., Freq. Control}, vol. 68, no. 9, pp. 2896--2905, Sep. 2021, doi: 10.1109/TUFFC.2021.3055138.

\bibitem{b24} P. Mar\'{e}chal, F. Levassort, L. -P. Tran-Huu-Hue, and M. Lethiecq, ``Lens-focused transducer modeling using an extended KLM model,'' \emph{Ultrasonics}, vol. 46, no. 2, pp. 155--167, May. 2007, doi: 10.1016/j.ultras.2007.01.006.

\bibitem{b25} H. S. Gougheri, A. Dangi, S. -R. Kothapalli, and M. Kiani, ``A comprehensive study of ultrasound transducer characteristics in microscopic ultrasound neuromodulation,'' \emph{IEEE Trans. Biomed. Circuits Syst.}, vol. 13, no. 5, pp. 835--847, Oct. 2019, doi: 10.1109/TBCAS.2019.2922027.

\bibitem{b26} E. Mer\v{c}ep, J. L. Herraiz, X. L. De\'{a}n-Ben, and D. Razansky, ``Transmission-reflection optoacoustic ultrasound (TROPUS) computed tomography of small animals,'' \emph{Light Sci. Appl.}, vol. 8, Jan. 2019, Art. no. 18, doi: 10.1038/s41377-019-0130-5.

\bibitem{b27} H. Deng, L. Gu, Y. Bai, C. Ma, and B. Luo, ``Photoacoustic/ultrasound dual-modality imaging of Sentinel lymph node with carbon nanoparticles,'' \emph{J. Phys.: Conf. Ser.}, vol. 2822, Sep. 2024, Art. no. 012037, doi: 10.1088/1742-6596/2822/1/012037.

\bibitem{b28} D. Kang, ``Signal magnitude nonlinearity to an absorption coefficient in photoacoustic imaging,'' \emph{J. Opt. Soc. Am. A}, vol. 37, no. 1, pp. 163--173, Jan. 2020, doi: 10.1364/JOSAA.37.000163.

\bibitem{b29} A. L\'{o}pez-Mar\'{i}n et al., ``Acoustic stack for combined intravascular ultrasound and photoacoustic imaging,'' \emph{IEEE Trans. Ultrason., Ferroelectr., Freq. Control}, vol. 72, no. 1, pp. 77--86, Jan. 2025, doi: 10.1109/TUFFC.2024.3465837.

\bibitem{b30} S. Cho et al., ``An ultrasensitive and broadband transparent ultrasound transducer for ultrasound and photoacoustic imaging in-vivo,'' \emph{Nat. Commun.}, vol. 15, Feb. 2024, Art. no. 1444, doi: 10.1038/s41467-024-45273-4.

\bibitem{b31} C. Qiu et al., ``Transparent ultrasonic transducers based on relaxor ferroelectric crystals for advanced photoacoustic imaging,'' \emph{Nat. Commun.}, vol. 15, Dec. 2024, Art. no. 10580, doi: 10.1038/s41467-024-55032-0.

\bibitem{b32} Z. Zhang, Z. Wang, S. Yang, F. Li, and Q. Ke, ``Textured ferroelectric ceramics based 1–3 piezoelectric composite for photoacoustic imaging'' \emph{Sens. Actuators A Phys.}, vol. 380, Dec. 2024, Art. no. 116030, doi: 10.1016/j.sna.2024.116030.

\bibitem{b33} Y. Cai, X. Luo, L. Xu, Z. Chen and J. Ma, ``Broadband stack-layer 3 MHz - 11 MHz dualfrequency ultrasound transducers for photoacoustic imaging,'' in \emph{Proc. IEEE Int. Ultrason. Symp. (IUS)}, Venice, Italy, 2022, pp. 1-3, doi: 10.1109/IUS54386.2022.9958818.

\bibitem{b34} W. Xia et al., ``An optimized ultrasound detector for photoacoustic breast tomography,'' \emph{Med. Phys.}, vol. 40, no. 3, Mar. 2013, Art. no. 032901, doi: 10.1118/1.4792462.

\bibitem{b35} X. Zhu et al., ``Real-time whole-brain imaging of hemodynamics and oxygenation at micro-vessel resolution with ultrafast wide-field photoacoustic microscopy,'' \emph{Light Sci. Appl.}, vol. 11, May. 2022, Art. no. 138, doi: 10.1038/s41377-022-00836-2.

\bibitem{b36} B. Lashkari and A. Mandelis, ``Linear frequency modulation photoacoustic radar: Optimal bandwidth and signal-to-noise ratio for frequency-domain imaging of turbid media,'' \emph{J. Acoust. Soc. Am.}, vol. 130, no. 3, pp. 1313--1324, Sep. 2011, doi: 10.1121/1.3605290.

\bibitem{b37} B. Lashkari and A. Mandelis, ``Time and frequency-domain biomedical photoacoustic imaging: A comparative study,'' in \emph{Proc. SPIE}, vol. 7883, \emph{Photonic Therapeutics and Diagnostics VII}, San Francisco, CA, USA, Feb. 2011, Art. no. 78834R, doi: 10.1117/12.880439.

\bibitem{b38} X. Jian, Z. Li, Z. Han, J. Xu, P. Liu, and Y. Liu, ``The study of cable effect on high-frequency ultrasound transducer performance,'' \emph{IEEE Sens. J.}, vol. 18, no. 13, pp. 5265--5271, Jul. 2018, doi: 10.1109/JSEN.2018.2838142.

\bibitem{b39} H. A. Kunkel, S. Locke, and B. Pikeroen, ``Finite-element analysis of vibrational modes in piezoelectric ceramic disks,'' \emph{IEEE Trans. Ultrason., Ferroelectr., Freq. Control}, vol. 37, no. 4, pp. 316--328, Jul. 1990, doi: 10.1109/58.56492.

\bibitem{b40} S. Sherrit and B.K. Mukherjee, ``Characterization of Piezoelectric Materials for Transducers,'' \emph{arXiv preprint}, arXiv:0711.2657 [cond-mat.mtrl-sci], Nov. 2007, doi: 10.48550/arXiv.0711.2657.

\bibitem{b41} IEEE Standard on Piezoelectricity, ANSI/IEEE Std 176-1987, IEEE, New York, NY, USA, 1988.

\bibitem{b42} E. P. EerNisse, ``Variational method for electroelastic vibration analysis,'' \emph{IEEE Trans. Sonics Ultrason.}, vol. 14, no. 4, pp. 153--159, Oct. 1967, doi: 10.1109/T-SU.1967.29431.

\bibitem{b43} K. Nakamura, ``Modelling ultrasonic-transducer performance: one-dimensional models,'' in \emph{Ultrasonic Transducers: Materials and Design for Sensors, Actuators and Medical Applications}, Sawston, Cambridge, UK: Woodhead Publishing Limited, 2012, ch. 6, sec. 6.3, pp. 196--198. \text{[Online]}. Available: \url{https://books.google.com.hk/books/about/Ultrasonic_Transducers.html?id=MZVwAgAAQBAJ&redir_esc=y}

\bibitem{b44} J. L. Butler and C. H. Sherman, ``Transducer models,'' in \emph{Transducer and Arrays for Underwater sound}, 2nd ed. Switzerland: Springer, 2016, ch. 3, sec. 3.1.1, p. 94. \text{[Online]}. Available: \url{https://link.springer.com/book/10.1007/978-3-319-39044-4}

\bibitem{b45} W. P. Mason, ``Electrical network theory,'' in \emph{Electromechanical Transducers and Wave Filters}, 2nd ed. New York, NY, USA: D.Van Nostrand Company, Inc. 1943, ch. 2, sec. 2.8, p. 60. \text{[Online]}. Available: \url{https://archive.org/details/in.ernet.dli.2015.13601}

\bibitem{b46} D. M. Pozar, ``Transmission line theory,'' in \emph{Microwave Engineering}, 4th ed. Hoboken, NJ, USA: John Wiley \& Sons, Inc. 2012, ch. 2, sec. 2.1, p. 48. \text{[Online]}. Available: \url{https://drive.google.com/file/d/0B42R6NUvjMKxTl9fM1c4a2JST28/edit?resourcekey=0-joVCaMFqbku8gE1IjD1GUA}


\end{thebibliography}
\end{document}